\documentclass[aps,prd,twocolumn,groupedaddress]{revtex4}

\usepackage{amsmath}
\usepackage{amssymb}
\usepackage{hyperref}
\usepackage{slashed}
\usepackage{graphics}
\usepackage{bm,bbm}
\usepackage{color}
\usepackage{enumerate}

\newcommand{\bdot}{{\boldsymbol{\cdot}}}

\newcommand{\hh}{{\hspace{.3mm}}}

\def\sideremark#1{\ifvmode\leavevmode\fi\vadjust{\vbox to0pt{\vss
 \hbox to 0pt{\hskip\hsize \hskip-0.5em
 \vbox{\hsize1cm\tiny\raggedright\pretolerance10000
  \noindent #1\hfill}\hss}\vbox to8pt{\vfil}\vss}}}

\newcommand{\rpl}                         
{\mbox{$
\begin{picture}(12.7,8)(-.5,-1)
\put(0,0.2){$+$}
\put(4.2,2.8){\oval(7.8,7.8)[r]}
\end{picture}$}}

\newcommand{\lpl}                         
{\mbox{$
\begin{picture}(12.7,8)(-.5,-1)
\put(1.66,-1.64){\scalebox{1.25}{$+$}}
\put(6.4,1.5){\oval(8,8)[l]}
\end{picture}$}}

\begin{document}


\title{Quantum Mechanics and Hidden Superconformal Symmetry}



\author{\href{https://www.youtube.com/watch?v=KaOC9danxNo}{R.~Bonezzi}}
\affiliation{
Groupe de M\'ecanique et Gravitation, Unit of Theoretical and Mathematical Physics, University of Mons (UMONS), 20 place du Parc, 7000 Mons, Belgium, {\tt roberto.bonezzi@umons.ac.be},
}

\author{O.~Corradini}
\affiliation{
 Dipartimento di Scienze Fisiche, Informatiche e Matematiche,\\ Universit\`a degli Studi di Modena e Reggio Emilia, Via Campi 213/A,  I-41125 Modena, Italy, \\and INFN, Sezione di Bologna, Via Irnerio 46, I-40126, Bologna, Italy,
 {\tt olindo.corradini@unimore.it}, }

\author{E.~Latini}
\affiliation{
Dipartimento di Matematica, Universit\`a di Bologna, Piazza di Porta S. Donato 5, I-40126 Bologna, Italy,  \\and INFN, Sezione di Bologna, Via Irnerio 46, I-40126, Bologna, Italy, {\tt emanuele.latini@UniBo.it},
 }

\author{A.~Waldron}
\affiliation{Department of Mathematics and the Center for Quantum Mathematics and Physics (QMAP), University of California, Davis, CA 95616, USA, {\tt wally@math.ucdavis.edu}.}


\date{\today}

\begin{abstract}
\noindent
Solvability of the  ubiquitous quantum harmonic oscillator relies on a spectrum generating~$\mathfrak{osp}(1|2)$ superconformal symmetry. 
We study the problem of constructing all quantum mechanical models with a hidden $\mathfrak{osp}(1|2)$ symmetry on a given  space of states. 
This problem stems from interacting higher spin models coupled to gravity. In one dimension, we show that the solution to this problem is the Plyushchay family of quantum mechanical models  with hidden superconformal symmetry
obtained 
by viewing
the harmonic oscillator as a one dimensional Dirac system, so that
Grassmann parity equals wavefunction parity.
These models---both oscillator and particle-like---realize all possible unitary irreducible representations of $\mathfrak{osp}(1|2)$.  \end{abstract}


\maketitle


\section{Introduction}\label{intro}

\noindent
The quantum harmonic oscillator 
$$H=  \frac12 \big( p^2 + q^2\big)\, ,\qquad [ p, q]=-i \, ,$$
is solvable because the ladder operators
\begin{equation}\label{ladder}
a=\frac{q+ip }{\sqrt{2}}\, ,\qquad a^\dagger=\frac{ q-ip }{\sqrt{2}}\, ,
\end{equation}
generate the spectrum.  This is perhaps the simplest example of the Lie superalgebra~$\mathfrak{osp}(1|2)$: to see this, one treats the ladder operators as supercharges~\footnote{Readers familiar with the the Dirac operator $\gamma^\mu \nabla_\mu$ playing the {\it r\^ole} of a SUSY charge, may wish to call $a:=\partial/\partial z$ and $a^\dagger:=z$ and then introduce a single Grassmann coordinate $\gamma$ obeying the Clifford algebra $\{\gamma,\gamma\}=2$. Then the SUSY charge $S^-=\gamma\,  \partial/\partial z$. In the above, $\gamma$ has been represented by the $1\times 1$ matrix $1$.}
$$S^+:=a^\dagger\, ,\qquad S^-:=a\, .$$  
Then defining the $\mathfrak{sp}(2)$ generators~\footnote{Indeed the space ${\mathcal B}$ of even number operator eigenstates of the harmonic oscillator form the metaplectic representation of $\mathfrak{sl}(2,{\mathbb R})\cong\mathfrak{sp}(2)$. 
The full Fock space has the ${\mathbb Z}_2$ graded decomposition ${\mathcal H}={\mathcal B}\oplus {\mathcal F}$.
}
$$Q^{++}=(a^\dagger)^2\, ,\quad Q^{+-}= H\, ,\quad Q^{--}=a^2\, ,$$
the five generators $\{S^\pm,Q^{\pm\pm},Q^{+-}\}$ generate the algebra~\footnote{Here we have suppressed vanishing relations and the~$\mathfrak{sp}(2)$ algebra obeyed by $\{Q^{++},Q^{+-},Q^{--}\}$.} $\mathfrak{osp}(1|2)$:
\begin{gather}
\{S^{\pm},S^{\pm}\}=2 Q^{\pm\pm}\, , \quad
\{S^+,S^-\}=2Q^{+-}\, ,\nonumber
\\
[S^{\mp},Q^{\pm\pm}]=\pm 2S^{\pm}\, , \quad  
[Q^{+-},S^{\pm}]=\pm S^{\pm}\, .  \label{osp}
\end{gather}
Strangely enough, here one assigns the ladder operators a Grassmann odd grading, even though these are the standard complex combinations of position and momentum given in Eq.~\eqref{ladder}. Thus, the fermion number operator $F$ that grades the $\frak{osp}(1|2)$ algebra counts one for odd powers of ladder operators (and zero for even powers) and therefore labels wavefunction parity~\footnote{Recall that harmonic oscillator eigenstates $|n\rangle$ are given by a Gaussian multiplied by Hermite polynomials, which are parity even (odd) when $n$ is even (odd).
In a coherent state picture, we could alternatively view the operator $a$ as a one dimensional Dirac operator $\gamma \frac{\partial }{\partial z}$ where $\gamma$ obeys the Clifford algebra $\{\gamma,\gamma\}=2$.}. 

The basic question we address is the existence of  operator quintuples  acting on the harmonic oscillator Fock space obeying the~$\mathfrak{osp}(1|2)$ Lie superalgebra. 
The solution to this problem is a class of quantum mechanical models that have been  studied in detail by Plyushchay~\cite{P1}.
We  also answer this question for generalized particle models with plane wave-normalizable spectra for which the $\mathfrak{osp}(1|2)$ algebra
acts  as a generalized one dimensional superconformal symmetry. 


Our study is motivated by a  proposal of Bars {\it et al}~\cite{BarsRey}, who suggested that the space of operators obeying an~$\mathfrak{sp}(2)$ algebra and acting on functions of a $d+2$ dimensional spacetime with two times, could describe gravitating, interacting higher spin theories. We have shown~\cite{BonezziO,Bonezzi} that this proposal is intimately linked to the study of~$d$ dimensional conformal geometries in terms of a $d+2$ dimensional ambient space initiated by Fefferman and Graham~\cite{FG}. The inclusion of fermions in such models leads to an $\mathfrak{osp}(1|2)$ generalization of Bars' theory~\cite{BonezziDirac} (see also~\cite{BarsReview}).
The study of the  detailed spectra, interactions, ultraviolet and unitarity properties of such models is a complicated problem commensurate with that  of string field theories, as one is dealing with field equations for operator-valued fields. 
Although the solution we find in one dimension is largely controlled by orthosymplectic representation theory, the existence of a mathematically well-defined answer in this setting is an important first step towards analyzing models in $d+2$ dimensions, for which the solution space already includes all~$d$-dimensional conformal geometries.
Moreover D'Hoker and Vinet~\cite{dH1}
have analyzed a hidden $\mathfrak{osp}(1|2)$ symmetry of the Dirac equation in monopole backgrounds,  which indicates tractability for models in higher dimensions.

Our analysis begins in Section~\ref{sec:EoM} with the ``master'' equations of motion and gauge symmetries for the supercharges $S^\pm$. Sections~\ref{sec:ops}-\ref{sec:VII}
are devoted to solving these equations on a harmonic oscillator Fock space while Sections~\ref{sec:VIII}-\ref{sec:X} focus on particle models with hidden superconformal symmetry. Appendix~\ref{repth} reviews $\mathfrak{osp}(1|2)$ representations.



\section{Equations of motion}
\label{sec:EoM}

\noindent
To answer the question posed in the introduction,
we view  the supercharges~$S^{\pm}$ as the 
fundamental ``fields'' and
 study ``equations of motion'' for these that guarantee that the algebra~$\mathfrak{osp}(1|2)$ of Eq.~\eqref{osp} holds. These have been formulated in~\cite{BonezziDirac}; the result is~\footnote{Note that these derive from a simple action principle~\cite{BonezziDirac} $S={\rm tr}\big[S^+S^-+\tfrac12 S^+ S^+ S^- S^-\big]$, where ${\rm tr}$ denotes an operator trace.}:
\begin{eqnarray}\label{eom}
[S^-, S^+S^+] &=&2S^+\, ,\nonumber\\[1mm]
[S^-S^-,S^+] &=&2S^-\, .
\end{eqnarray}
The statement here is that if the pair of operators~$S^\pm$ obey these equations, then the operator quintuple $\big\{S^\pm,\,  Q^{\pm\pm}=S^\pm S^\pm,\,  Q^{+-}=\tfrac12 [S^+S^-+S^+ S^-]\big\}$ satisfies the 
$\mathfrak{osp}(1|2)$ Lie superalgebra~\eqref{osp}.

Clearly, if $S^\pm$ solve Eqs.~\eqref{eom}, then so too do~$U^{-1} S^\pm U$ for any invertible operator $U$. Linearizing $U$ around the identity $U\approx {\rm Id}+ \varepsilon$ gives the  gauge invariance
\begin{equation}\label{gauge}
S^\pm \sim S^\pm + [S^\pm,\varepsilon]\, ,
\end{equation}
of the equations of motion~\eqref{eom}. Here the gauge parameter~$\varepsilon$ is itself also an operator. 

The problem of solving Eqs.~\eqref{eom} for operators $S^{\pm}$ is not defined without specifying the state space~${\mathcal H}$ on which these operators act~\footnote{Alternatively, one may first solve for a set of operators on some space and only thereafter search for an appropriate inner product.}. 
The set of  possible choices for an
underlying Hilbert space~${\mathcal H}$ is clearly enormous.
We commence with
perhaps the simplest case in which~${\mathcal H}$ is the harmonic oscillator Fock space.

\section{The Space of Operators}
\label{sec:ops}

\noindent
We now let 
${\mathcal H}$ equal the quantum harmonic oscillator Hilbert space with  Fock basis $\{|n\rangle \,\!\! : \,\!\! n\in {\mathbb Z}_{\geq 0}\}$. We employ the slightly non-standard normalization $\langle m|n\rangle= n! \, \delta_{mn}$ for states~$|n\rangle$, since this allows us to identify $|n\rangle$ with the monomial $z^n$ and in turn study  wavefunctions given by polynomials, or more generally suitable analytic functions, in $z$~\footnote{Put simply, as a convenient bookkeeping device, we identify $a^\dagger \leftrightarrow z$, $a \leftrightarrow \tfrac{\partial}{\partial z}$ and $|0\rangle\leftrightarrow 1$.}. Thus we study operators 
\begin{equation}\label{ansatz}
S^{\pm}=s_0^{\pm}(z) + s_1^{\pm}(z)\frac{\partial}{\partial z}
+ s_2^{\pm}(z)\frac{\partial^2}{\partial z^2}+\cdots\, ,
\end{equation}
where $s_i^\pm(z)$ are analytic functions of $z$ in a neighborhood of the origin. In terms of  ladder operators,  this amounts to studying operators given by sums of normal ordered products of $a$'s and $a^\dagger$'s. More precisely, we are looking for  the most general set of formal power series in ladder operators obeying the~$\mathfrak{osp}(1|2)$ superalgebra. 

\section{Gauge choices}
\label{sec:gauge}
\noindent
To simplify our problem we  fix a gauge using the freedom in Equation~\eqref{gauge}. A propitious choice~is 
\begin{equation}\label{z}
S^+=z\, .
\end{equation}
To verify gauge  reachability, we consider
$$
\varepsilon = \epsilon_0(z) + \epsilon_1(z)\frac{\partial}{\partial z}
+ \epsilon_2(z)\frac{\partial^2}{\partial z^2}+\cdots\, .
$$
Then a short computation gives
$$
[\varepsilon,z]=\epsilon_1(z) + 2\epsilon_2(z) \frac{\partial}{\partial z}
+
3\epsilon_3(z) \frac{\partial^2}{\partial z^2}+\cdots\, .
$$
Thus by solving for $\epsilon_1(z)$, $\epsilon_2(z)$,... we can bring $S^+=z$ to an operator of the general  form~\eqref{ansatz} by a gauge transformation~\eqref{gauge}.
The function $\epsilon_0(z)$ remains undetermined because
there are still residual gauge transformations, respecting our choice $S^+=z$, of the form $S^\pm\mapsto \big(1/U(z)\big)\,  S^{\pm} \hh U(z)$. 
The beauty of the gauge choice~\eqref{z} is that the first equation of motion in~\eqref{eom} is now linear.

\section{The linear equation}
\label{sec:lineq}

\noindent
The  linear equation for  $S^-$ reads
\begin{equation}\label{first}
[S^-,z^2]=2z\, .
\end{equation}
Using the identity
$$
\Big[\frac{\partial^k}{\partial z^k},z^2\Big]=k\Big[2z\frac{\partial}{\partial z}+(k-1)
\Big]\frac{\partial^{k-2}}{\partial z^{k-2}}\, ,
$$
we can solve this order by order for $S^-$ and find
$$
S^-=\frac{\partial}{\partial z} + A(z) 
+ B(z) \Big[
1- z\frac{\partial}{\partial z} + \frac{2}{3} z^2 
\frac{\partial^2}{\partial z^2}+\cdots
\Big]\frac{\partial}{\partial z}\, .
$$
In the above $A(z)$ and $B(z)$ are arbitrary functions.
Defining the number operator $N:=z\tfrac{\partial}{\partial z}$, and denoting normal ordering by $\colon\!\bullet\hh \hh\colon$ ({\it e.g.},  $\colon\!N^2\colon =z^2\tfrac{\partial^2}{\partial z^2}=N(N-1)$), the above display becomes~\footnote{Observe that $[1-\exp(-2N)]/N$ is analytic in $N$.} 
$$
S^-=\frac{\partial}{\partial z} + A(z) 
+ B(z) \, \colon\!\!\Big[
\frac{1-e^{-2N}}{2N}
\Big]\!
\colon\, \frac{\partial}{\partial z}\, .
$$
Using the identity 
$$
z\, \colon \!f(N)\colon\frac{\partial}{\partial z}=\, \colon\! Nf(N)\colon\, ,
$$
we have
\begin{equation}\label{Sm}
S^-=\frac{\partial}{\partial z} + A(z) 
+ \frac{B(z)}z \, \colon\!\!\Big[
\frac{1-e^{-2N}}{2}
\Big]\!
\colon\, .
\end{equation}
The normal ordered operator in the above expression is related to the Klein operator of~\cite{P3}. It
has an interesting action on number operator eigenstates
$$
 \colon\!\!\Big[
\frac{1-e^{-2N}}{2}
\Big]\!
\colon |n\rangle = 
\frac 12 (1-(-1)^n) |n\rangle\, ,
$$
{\it i.e.,} it vanishes on the space of even number operator eigenstates~${\mathcal B}$ 
and is unity on the space of odd number operator eigenstates~${\mathcal F}$. 
This means that the operator~$1/z$ appearing in Eq.~\eqref{Sm} is well-defined.
Also, the operator in the above display  is the fermion number operator
$$
F\colon=\frac 12 (1-(-1)^N)=F^2\, .
$$
This obeys $\{F,z\}=z$ and $[F,z^2]=0$.
In addition to providing a ${\mathbb Z}_2$ grading of the Hilbert space
$$
{\mathcal H}={\mathcal B}\oplus{\mathcal F}\, ,
$$
we may demand that $F$ also coincides with the ${\mathbb Z}_2$ grading of the Lie superalgebra $\mathfrak{osp}(1|2)=\mathfrak{sp}(2)\oplus {\mathbb R}^2$. In the following we focus on the case where the two gradings coincide, since it leads quickly to the solution space; we prove that this yields the most general solution in Appendix~\ref{E's favorite appendix}.

\section{Harmonic Oscillator Solution}
\label{sec:VI}

\noindent
Requiring coincidence of ${\mathbb Z}_2$
gradings in conjunction with the solution to the linear equation~\eqref{Sm} forces us to consider an ansatz of the form
\begin{eqnarray}\label{ansatz1}
S^+&=& \:\: \, z\: ,\nonumber\\[1mm]
S^-&=&\frac\partial{\partial z}+ \alpha(z) + (-1)^F \beta(z)\, ,
\end{eqnarray}
where $\alpha(z)$ and $\beta(z)$ are both odd with respect to the ${\mathbb Z}_2$ grading ({\it i.e.}, even and odd functions of $z$). 
Here we also used that $(-1)^F=1-2F$.

It remains to solve the second, non-linear equation in~\eqref{eom} which we rewrite as
$$
[H,S^-]+S^-=0\,  ,
$$
where  the Hamiltonian is easily computed from
Eq.~\eqref{ansatz1}: 
$$
H=\frac12 \{S^+,S^-\} = N+\frac12 + z \alpha(z)\, .
$$
The above leads to the relation
$$
z\beta'(z)+\beta(z)=0\: \Rightarrow \beta(z)=\frac{c}{2z}\, ,
$$
for some constant $c$. Requiring that $S^-$ acting on the Fock space~${\mathcal H}$ (and in particular on the vacuum $|0\rangle$) is well-defined  we set
$$
\alpha(z)=\frac{c}{2z}+ A(z)\, ,
$$
where $A(z)$ is analytic and odd. Thus
\begin{equation}\label{abelian}
S^-=
\frac\partial{\partial z}+ A(z) + \frac{c}{z}\, F\, .
\end{equation}
First observe that since $F|0\rangle=0$, the operator $\frac{1}{z} F$ is, as promised, well-defined. Moreover,
since $A(z)$ is odd, the function $U(z)$~$=$~$\exp\big(-\int^z A(z)\big)$ is even and thus commutes with $F$. Hence $(1/U(z))\hh S^- U(z)= \tfrac{\partial}{\partial z}+\tfrac{c}z F$. 

The constant $(c+1)/2$ measures the zero point energy~$E_0$ of the vacuum $|0\rangle$, so we now call $c=2E_0-1$.
Altogether then, we find a one parameter family of solutions 
\begin{gather}
S^+=z ,\quad S^-=\frac\partial{\partial z}+ \frac{2E_0-1}{z}\, F ,\label{solution}\\[2mm]
Q^{+-}\!=N+E_0\, , \nonumber\\[1mm]
Q^{++}\!=z^2 ,\quad
Q^{--}\!=\!\frac{\partial^2}{\partial z^2}+\frac{2E_0-1}z
\frac{\partial}{\partial z}
-\frac{2E_0-1}{z^2}\, F\, .\nonumber
\end{gather}
Although the Hamiltonian $H=Q^{+-}$ only receives a shift in its zero point energy.
The commutator of the deformed oscillators $S^\pm$ is easily calculated to be
\begin{equation}\label{anticomm}
[S^-,S^+]=
 1-(2E_0-1)(2F-1)\, .
\end{equation}
This is exactly the model proposed by Plyushchay~\cite{P1} (although basic quantum commutators were already studied in~\cite{Wigner}).
The operator $S^-$ is a Yang--Dunkl type operator~\cite{Yang-Dunkl}.
The~$\mathfrak{osp}(1|2)$ representation obeyed by this model was analyzed in~\cite{P3} (and also recently discussed in~\cite{TheTop}), this is summarized in the next section~\footnote{The material presented in Section~\ref{sec:VII} was actually developed independently of~\cite{P3} and reproduces the results found there. The results on the singular vector are new.}.

\section{Oscillator orthosymplectic representation}
\label{sec:VII}

Our solution~\eqref{solution} obeys the $\mathfrak{osp}(1|2)$ Lie superalgebra
and therefore provides a representation thereof.
To analyze this we start by searching for states annihilated by~$S^-$ so consider
 $\psi(z)$ subject to
$$
S^-\psi(z)=0\, ,
$$
which we decompose as
$$
\psi(z)=\psi_+(z)+\psi_-(z)\, ,
$$
where the two terms on the right hand side are analytic and even/odd respectively. Since $S^-$ is odd we must separately have
$$
\left\{
\begin{array}{l}
\psi'_+(z)=0\, ,\\[2mm]
\psi'_-(z)+{\displaystyle\frac{2E_0-1}z} \, \psi_-(z)=0\, .
\end{array}
\right.
$$
Thus $\psi_+(z)=1=|0\rangle$, the standard Fock vacuum. There is in addition the possibility of a second solution 
$\psi_-(z)=z^{1-2E_0}$. Because $\psi_-(z)$ is analytic and odd this occurs only when $E_0=-n$ with 
$n\in {\mathbb Z}_{\geq 0}$, whence $\psi_-(z)=|2n+1\rangle=S_+^{2n+1}|0\rangle$.
Thus
$$
\ker S^-=\left\{
\begin{array}{ll}
{\rm span}\big\{|0\rangle\, , \: |2n+1\rangle
\big\}\, , & E_0\in {\mathbb Z}_{\leq 0}\, ,\\[3mm]
{\rm span}\big\{|0\rangle\big\}\, , &E_0\notin
 {\mathbb Z}_{\leq 0}\, .
 \end{array}
 \right.
$$
Thus $|0\rangle$ is always a highest weight state subject to
$$
H|0\rangle = E_0 |0\rangle\, ,
$$
while $|2n+1\rangle$ is a singular vector when $E_0=-n$ and then obeys
$$
H|2n+1\rangle=(n+1) |2n+1\rangle\, .
$$

At the harmonic oscillator value $E_0=\frac12$, we have $S^-=\partial/\partial z=a=(S^+)^\dagger$
and  $Q^{--}=(Q^{++})^\dagger$. The Hilbert space is then the unitary irreducible representation ${\mathcal S}(1/2)={\mathcal D}(1/2)\oplus
{\mathcal D}(3/2)$ given by a direct sum of two discrete series unitary irreducible  $\mathfrak{sp}(2)$ representations.  Indeed, unlike $\mathfrak{sp}(2)$, which also has supplementary and principal series representations, the Lie superalgebra $\mathfrak{osp}(1|2)$ only has discrete series unitary irreducible representations~\cite{Furutsu} (see Appendix~\ref{repth} for further details).

When $E_0\notin{\mathbb Z}_{\leq0}$, the even and odd states
${\mathcal B}=\{|0\rangle,|2\rangle,|4\rangle,\ldots\}$ and ${\mathcal F}=\{|1\rangle,|3\rangle,|5\rangle\ldots \}$, respectively,
separately diagonalize the $\mathfrak{sp}(2)$ Casimir
$$
c_{\mathfrak{sp}(2)}=\frac14\hh(Q^{+-})^2-\frac18 \hh \{Q^{++},Q^{--}\}
$$
which takes values
$$
c_{\mathfrak{sp}(2)}({\mathcal B})=\tfrac{E_0(E_0-2)}4\, \mbox{ and }\,
c_{\mathfrak{sp}(2)}({\mathcal F})=\tfrac{(E_0-1)(E_0+1)}4\, .
$$
When $E_0>0$, these precisely match the Casimirs of the discrete series representations ${\mathcal D}(E_0)$ and ${\mathcal D}(E_0+1)$.
Moreover, the direct sum of these representations yields the $\mathfrak{osp}(1|2)$ discrete series representation ${\mathcal S}(E_0)$.
Indeed, the orthosymplectic
Casimir 
$$
\hh c_{\mathfrak{osp}(1|2)}
=\frac14\hh(Q^{+-})^2-\frac18 \hh \{Q^{++},Q^{--}\}-\frac18\hh [S^+,S^-]\, ,
$$ 
obeys
\begin{equation*}
\hh c_{\mathfrak{osp}(1|2)}({\mathcal H})=\frac{E_0(E_0-1)}4=\hh c_{\mathfrak{osp}(1|2)}\big({\mathcal S}(E_0)\big)
\end{equation*}
on the harmonic oscillator state space ${\mathcal H}={\mathcal B}\oplus {\mathcal F}$.
However,
when $E_0\neq 1/2$, the operators $Q^{--}$ and $S^-$ are no longer the hermitean conjugates of $Q^{++}$ and $S^+$ with respect to the standard Fock space inner product.
But, since the $\mathfrak{osp}(1|2)$ action on
the harmonic oscillator Fock space is isomorphic to that of the orthosymplectic discrete series, there exists a corresponding inner product with respect to which this is a unitary representation. 
This inner product can be computed
as follows:

First observe that with respect to the Fock norm 
the state $|E_0,n\rangle=(S^+)^n|0\rangle=|n\rangle$ obeys
$$
\big|\!\hh\big||E_0,n\rangle\big|\!\hh\big|^2_{\rm Fock}=\langle 0 | a^n (a^\dagger)^n|0\rangle=n!\langle 0|0\rangle = n!\, .
$$
However, with respect to the unitary discrete series norm, 
\begin{equation*}
\begin{split}
\big|\!\hh\big||E_0,n\rangle\big|\!\hh\big|_{\mathfrak{osp}}^2
&=\big\langle |E_0,n\rangle,|E_0,n\rangle\big\rangle_{\mathfrak{osp}}
\\[1mm]
&\!\!\!
=\langle E_0,0|(S^-)^n(S^+)^n| E_0,0\rangle
\\[1mm]
&\!\!\!=\langle E_0,0|(S^-)^{n-1}S^-| E_0,n\rangle
\\[1mm]
&\!\!\!=
\big((E_0-\tfrac12)(1-(-1)^n)+n\big)
\big|\!\hh\big||E_0,n-1\rangle\big|\!\hh\big|_{\mathfrak{osp}}^2
\\[1mm]
&\!\!\!=
2^n(E_0)_{[\frac {n+1}2]} \big(\big[\tfrac {n}2\big]\big)\raisebox{-.5mm}{\scalebox{1.2}{$!$}}
\, .
\end{split}
\end{equation*}
Here we have employed the standard Pochhammer notation and  used the identity  (valid for $n\in {\mathbb Z}_{\geq 1}$)
\begin{eqnarray}
S^-|E_0,n\rangle=
\big((2E_0-1)(1-F)+n\big)
|E_0,n-1\rangle\, .\label{theid}
\end{eqnarray}
The operator version of this identity is given in~\eqref{anticomm}.

Importantly, the above derivation uses only the $\mathfrak{osp}(1|2)$ algebra. Hence we have the relation between Fock and discrete series inner products~\footnote{This relation was first uncovered in~\cite{P2}.} for the complete set of states $\{|E_0,n\rangle\, | n\in {\mathbb Z}_{\geq 0}\}$ 
\begin{multline*}
\big\langle | E_0,n\rangle,|E_0,m\rangle\big\rangle_{\mathfrak{osp}}
=
\frac{2^{[\frac{n+1}2]}(E_0)_{[\frac {n+1}2]}}
{(2\hh [\frac{n+1}2]-1)!!}
\langle E_0,n|E_0,m\rangle\quad
\\[1mm]
\quad= 2E_0n!\hh\delta_{n,m}
\big(1+\tfrac{2E_0-1}3\big)\big(1+\tfrac{2E_0-1}5\big)\cdots\big(1+\tfrac{2E_0-1}{2[\frac {n+1}2]-1}\big)
\, .
\end{multline*}
We would like to encode this using an operator built from the Casimir and number operators, and therefore note that 
\begin{equation}\label{CAZ}
\sqrt{4c_{\mathfrak{osp}(1|2)}+\tfrac14}\: \big|E_0,n\big\rangle=|E_0-\tfrac12|\, \big|E_0,n\big\rangle\, .
\end{equation}
Thus, by virtue of the identity~\eqref{theid}, we introduce the operator
$$
{\cal I}=
\frac{2^{\frac{N+F}2}\hh (\hat E_0)_{\frac{N+F}2}}
{(N+F-1)!!}\, ,
$$
where the operator-valued Pochhammer and double factorial are defined by expanding in eigenstates of $N$, while the operator  $\hat E_0$
returns $E_0$ on all states and can  be expressed in terms of the Casimir via~\eqref{CAZ}. 
By construction
$
 a\hh  {\mathcal I}|E_0,n\rangle = {\mathcal I} S^- |E_0,n\rangle$ whence
 $$
 a \hh{\cal I}={\cal I}\hh  S^-\, .
$$
Thus, the discrete series unitary inner product $\langle\bdot\hh,\bdot\rangle_{\mathfrak{osp}}$ between states $\Psi=|\psi\rangle$ and $\Phi=|\phi\rangle$ then reads
$$
\langle \Psi,\Phi\rangle_{\mathfrak{osp}} = \langle \psi | \hh {\cal I}\hh | \phi\rangle\, .
$$
Hence, when $E_0>0$ we have found a realization of the unitary orthosymplectic discrete series representations~${\mathcal S}(E_0)$ in terms of the harmonic oscillator state space.

\medskip

Finally, note that when $E_0=-n\in Z_{\leq 0}$ the harmonic 
oscillator no longer gives an irreducible orthosymplectic representation. However, the space of descendants ${\mathcal H}_-$ of the singular vector 
$$
|n+1,0\rangle := |2n+1\rangle\, ,\mbox{ where }
H|n+1,0\rangle = (n+1)|n+1,0\rangle \, ,
$$
form a unitary discrete series representation ${\mathcal S}(n+1)$ (with respect to the $E_0=n+1$ inner product).
The quotient ${\mathcal H}/{\mathcal H}_-$ then gives a finite dimensional (non-unitary)
orthosymplectic representation.

\section{Superconformal Quantum Mechanics}
~\label{sec:VIII}

\noindent
We now want to repurpose our harmonic oscillator analysis for a study of novel superconformal theories.
For that we will modify our Hilbert space such that 
the operator~$-\frac12 Q^{--}$ is self-adjoint and plays the {\it r\^ole} of the Hamiltonian~$H$. 
 We may then view $\mathfrak{osp}(1|2)$ as a conformal superalgebra: 
\begin{gather}
H=-\frac12(S^-)^2\,  ,\quad D=\frac12 \{S^+,S^-\}\, ,\quad K= \frac12(S^+)^2\, ,\nonumber\\
\quad iQ=S^-\, ,\quad  S=S^+\, .\nonumber
\end{gather}
Here, because $\mathfrak{osp}(1|2)$ imposes
$$
Q^2=2H\, ,
$$
the operator $Q$ is the SUSY generator. Also $D$ and $K$ correspond to dilations and conformal boosts while $S$ is the conformal SUSY charge.

We now need to build the Hilbert space on which $H$ and $Q$ act. For that we begin by studying the space of wavefunctions $\psi(x)$ on the line~${\mathbb R}$. Since the de Rham cohomology of this space is trivial, we will {\it assume} that the abelian gauge field $A$ appearing in Eq.~\eqref{abelian} can be gauged away in the following analysis. Thus the SUSY charge is 
$$
iQ
=\frac{\partial}{\partial x} + \frac{2E_0-1}{x}\, F\, .
$$
while half its square gives the Hamiltonian
$$
H=-\frac12\frac{\partial^2}{\partial x^2} -
\big(E_0-\tfrac12\big)\Big(
\frac 1{x}\frac{\partial}{\partial x} -
\frac{1}{x^2}\, F\Big)\, .
$$
In the above displays, the fermion occupation number $F$ equals unity on odd wavefunctions $\psi_-(x)=-\psi_-(-x)$ and vanishes on even wavefunctions 
$\psi_+(x)=\psi_+(-x)$. The remaining $\mathfrak{osp}(1|2)$ generators are obtained by the replacement $z\mapsto x$ in the solution given in Eq.~\eqref{solution}. Observe, that the ${\mathbb Z}_2$ grading $\mathfrak{osp}(1|2)={\mathcal B}\oplus {\mathcal F}$ with ${\mathcal B}={\rm span}\{Q^{\pm\pm},Q^{+-}\}$ and ${\mathcal F}={\rm span}\{S^{\pm}\}$ still holds when $F$ is defined by wavefunction parity.

The inverse square potential in the above Hamiltonian is typical of conformal quantum mechanical models~\cite{dAFF}.  Supersymmetry charges and Hamiltonians of this type were also studied  by Plyushchay~\cite{P1,P2,P3}. Our next task is to develop an inner product with respect to which they are self-adjoint.
This will require a careful analysis of the space of self-adjoint extensions for these operators~\footnote{A similar analysis for the case of the Dirac operator in monopole backgrounds has been performed in~\cite{dH2}.}. 
 
\section{The Inner Product}

\noindent
Our first task is to ensure definite hermiticity for the supercharge $Q$ (thereafter we will examine its self-adjointness).
 For that, first observe that acting on odd functions $iQ$ simply acts as $\frac{\partial}{\partial x} + \frac{2E_0-1}{x}$. Therefore it is convenient to define
\begin{align}\label{oddparam}
\psi_-(x) =: x^{1-2E_0} \tilde \psi(x)  
\end{align}
so that we have the identity
\begin{align*}
iQ\psi_-(x) =x^{1-2E_0} \frac{\partial}{\partial x}\tilde \psi(x)~.
\end{align*} 
Note that $E_0$ is an, {\it a priori} arbitrary, complex number. Firstly let decompose wavefunctions into even and odd parts according to
$$
\psi=\psi_++\psi_-\, ,
$$
and then use that the information of $\psi$ is stored by $\psi_\pm$ on the positive half line $x>0$. On the whole line we thus define~\footnote{{\it A posteriori} we will check that the behavior at $x=0$ is smooth, {\it i.e.}, no powers of $|x|$ are involved.}
\begin{align*}
 \psi_\pm(x) :=\left\{\begin{array}{ll}
\,\, \,\,   \psi_\pm(x)\,,& x>0\, ,\\[1mm]
\pm\psi_\pm(-x)\,,& x<0\, .
\end{array}\right.
\end{align*}
 Using the parametrization~\eqref{oddparam} for the odd part  we may thus define the inner product
\begin{align}\label{eq:inner}
\langle\varphi,\psi\rangle:&=2\int_0^\infty dx\, x^{2E_0-1}\Big[  \varphi_+^*\psi_+ +\, \varphi_-^*\psi_-\Big]\nonumber\\
&=2\int_0^\infty dx \Big[x^{2E_0-1}\,  \varphi_+^*\psi_++x^{1-2E_0}\, \tilde\varphi^*\tilde \psi\, \Big]
\, .  
\end{align}
 For $E_0\in {\mathbb R}$, this inner product is positive definite and sesquilinear, but restricts the allowed behavior of wavefunctions at $x=0,\infty$. In particular $\psi_\pm$ must be square integrable with respect to the measure $x^{2E_0-1}$ on~${\mathbb R}$.
In particular this requires that for small $x$, the fastest decay behavior of $\psi_\pm$ is $\psi_\pm\sim x^{a_\pm}$, with 
\begin{align}
&a_\pm >-E_0\, .
\label{eq:H}
\end{align}
We denote the space of functions with square integrable behavior at large $x$ and decay rate at the origin satisfying the above bound by  ${\mathcal H}_{a_+,a_-}$.  We next examine the SUSY charge on these spaces.

Now since
$
iQ\psi = \psi_+'\, +\,x^{1-2E_0}\tilde \psi'$ (primes denote $x$ derivatives),
it follows that
$$(iQ\psi)_+=x^{1-2E_0} \tilde\psi'\, ,\,\, \, 
(iQ\psi)_-=\psi'_+\, .
$$
A wavefunction $\psi$ sits inside the domain ${\rm dom}(Q)$ of $Q$ provided it has the following small-$x$ behavior
\begin{align}
\psi_\pm \sim x^{a_\pm}\, ,\quad
 a_\pm >1-E_0\, .\label{eq:domQ}
\end{align}
The operator $Q$ is hermitian, since 
\begin{eqnarray}
&& \!\!\!\langle \theta, Q\psi\rangle^*=
2i\int_{0}^\infty dx \big[\theta_+^*\tilde\psi' + \tilde \theta^*\psi'_+\big]^*\nonumber\\
&=& -2i\int_{0}^\infty dx \big[\psi_+^*\tilde\theta' + \tilde \psi^*\theta'_+\big] -2i\big[ \psi_+^*\tilde\theta + \tilde \psi^*\theta_+\big]|_{x=0}\nonumber \\
&=& \langle \psi, Q\hh \theta\rangle\,,\quad \forall \hh\theta,\, \psi \in {\rm dom}(Q)\, .\label{parts}
\end{eqnarray}
In the above, 
the condition~\eqref{eq:domQ} guarantees  cancellation of the boundary term, which only requires the (weaker) condition $a_+ + a_- > 1-2E_0$.
Thus the SUSY charge is hermitean (indeed we chose the inner product~\eqref{eq:inner} precisely for this reason). It remains to examine whether~$Q$ is (essentially) self-adjoint, or more precisely whether it admits self-adjoint extensions. The following analysis is standard and follows classical work by Von Neumann~\cite{Reed}. 
 
 The space ${\rm dom}(Q)$ is dense in ${\mathcal H}$ so $Q$ possibly has self-adjoint extensions. The dimension of the space of extensions equals 
 the dimensions of $[{\rm ran}(Q\pm i\lambda)]^\perp$ for $\lambda$ real and positive---if these dimensions differ for $\pm\lambda$ the operator $Q$ has no self-adjoint extensions---these dimensions are known as {\it deficiency indices}. It is, of course, equivalent to compute the dimensions of 
 $\ker(Q\pm i\lambda)$ and
 the  condition $Q\psi=-i\lambda \psi$ amounts to 
\begin{equation}\label{kernel}
\psi_+'=\lambda \psi_-\, ,\quad x^{1-2E_0}\big(x^{2E_0-1} \psi_-\big)' = \lambda \psi_+\, .
\end{equation}
These  can be reduced to a pair of modified Bessel equations: 
We call $y=\lambda x$ and $\psi_\pm(x) = x^{1-E_0} u_\pm(y)$ and feed the two equations into one another which gives 
\begin{equation}\label{Bessel}
u_{\pm}''(y) + \frac{1}{y} u_\pm'(y) - \Bigg[1+\frac{\alpha_\pm^2}{y^2}\Bigg]u_\pm(y)=0\, ,
\end{equation}
where
$\alpha_+=E_0-1$ and $\alpha_-=E_0$.

Equations~\eqref{Bessel} are identical for both $\pm\lambda$, so that the deficiency indices are equal. Solutions to~\eqref{Bessel} are modified Bessel functions ($I_\alpha,\, K_\alpha$) with indices~$\alpha_\pm$. Of these solutions only $K_\alpha(\lambda x)$ has a good behavior at $x\to \infty$. On the other hand, for small, positive, $x$ it behaves (up to a non-zero coefficient) as~\footnote{For $\alpha =0$, $K_0$ behaves logarithmically and the corresponding wave functions are not normalizable.}
\begin{align*}
K_\alpha (\lambda x) \sim x^{-|\alpha|}\, ,
\end{align*}  
so that
\begin{align*}
\psi_+(x)\sim x^{1-E_0-|E_0-1|}\,,\quad \psi_-(x)\sim x^{
1-E_0-|E_0|}\, .
\end{align*}
Hence, 
solutions to the kernel condition~\eqref{kernel}
  are in~${\mathcal H}$ if the above exponents satisfy the condition~\eqref{eq:H}, which amounts to
\begin{align}
0<E_0<1\, .
\label{eq:c}
\end{align} 
In other words, when the parameter $E_0$ satisfies the above condition
both deficiency indices are unity and there is a one-parameter set of self-adjoint extensions~\footnote{The problem 
of finding the boundary conditions for operators of the form $-\frac{\partial^2}{\partial x^2}+(\alpha-\frac14)\frac1{x^2}$ on the half-line is of topical interest in the analysis literature, see~\cite{Derezinski}
}.
 On the other hand if $E_0$ does not satisfy~\eqref{eq:c} there is a unique extension. Since $2H=Q^2$, it follows that the Hamiltonian also has a unique self-adjoint extension in the latter case. Moreover, we immediately learn that  the spectrum of $H$ is bounded below by zero. 
This can also be seen by explicitly computing the expectation value of the Hamiltonian for some state $\psi=\psi_++\psi_-:=\psi_++x^{\frac12-E_0} \chi\hh$:
\begin{widetext}
\begin{eqnarray*}
\langle \psi, H \psi\rangle&=&
-\int_0^\infty dx\hh x^{2E_0-1} \left[\psi_+^*\psi''_++\frac{2E_0-1}x \hh
\psi_+^* \psi_+'\right]
-\int_0^\infty dx\hh x^{2E_0-1} \left[\psi_-^*\psi''_-+\frac{2E_0-1}x \hh
\Big(\psi_-^* \psi_-'-\frac{|\psi_-|^2}x\Big)
\right]\\[2mm]
&=&
\int_0^\infty dx\hh x^{2E_0-1} |\psi'_+|^2\, 
+\int_0^\infty
dx\hh
\left[|\chi'|^2+\frac{E_0^2-\frac14}{x^2}\, |\chi|^2\right]
\hh=\hh
\int_0^\infty dx\hh x^{2E_0-1} |\psi'_+|^2\, 
+\int_0^\infty
dx\hh
\Big|
\chi'+\frac{E_0-\frac12}x \hh \chi
\Big|^2\, .
\end{eqnarray*}
\end{widetext}
Here we have used that $\psi$ is in the domain of $H$ to kill boundary terms at the origin generated by integrations by parts in the above computation. The final result is manifestly positive for all $E_0$ (even though the Hamiltonian has non-positive potential term for $E_0<\frac12$ acting on odd wavefunctions).



\vfill\eject

\section{The Spectrum}
~\label{sec:X}

\noindent
To compute the spectrum of the model we diagonalize the SUSY charge~$Q$
in order to solve the Schr\"odinger equation $H\psi=E\psi$.
The ``BPS'' states obeying $Q\psi=0$ are constants which are not finite norm.
This indicates that we expect to find a plane wave normalizable spectrum, just as for the free particle on a line.

Indeed, we may recycle our deficiency index computation to  solve $H\psi=E\psi$ by replacing $\lambda\to i\sqrt{2E}$. 
We find $\psi_\pm=x^{1-E_0} v_{\pm}(\sqrt{2E}\hh x)$ where $v_\pm(y)$ obeys the Bessel equation
$$
v_{\pm}''(y) + \frac{1}{y} v_\pm'(y) + \Bigg[1-\frac{\nu_\pm^2}{y^2}\Bigg]v_\pm(y)=0\, ,
$$
with indexes
\begin{equation}\label{inds}
\nu_+=|E_0-1|\, , \quad \nu_-=|E_0|\, .
\end{equation}
Here we have chosen $\nu_\pm\geq 0$ in order that we get plane wave normalizable solutions.
Thus we have 
$$
\psi_E(x)=\frac{\beta_+ \, 
 J_{|E_0-1|}(\sqrt{2E}x)
+
\beta_- 
\, J_{|E_0|}(\sqrt{2E}x)}{x^{E_0-1}}\, ,
$$
where the complex constants $\beta_\pm$  multiply the even/odd solutions. It follows from our previous deficiency index computations that these solutions are not normalizable, nonetheless, they obey an analog of plane wave normalizability by virtue of the closure relation for Bessel functions (valid for $\nu>-1/2$ and hence for any values of the positive indexes $\nu_\pm$ in Eq.~\eqref{inds})
\begin{align*}
\int_0^\infty x dx\,  J_\nu(\sqrt{2E} x) J_\nu(\sqrt{2E'}x)&=\frac{\delta(\sqrt{2E}-\sqrt{2E'})}{\sqrt{2E}}\\[1mm]
&=\delta(E-E')\, .
\end{align*}
Indeed, if we define Bose and Fermi scattering states by
$$\, 
|E,+\rangle=\frac{ J_{|E_0-1|}(\sqrt {2E} \hh x)}{ \sqrt2\, x^{E_0-1}}\, ,\quad
|E,-\rangle=\frac{J_{|E_0|}( \sqrt {2E}\hh x)}{\sqrt2\,x^{E_0-1}}\, ,
$$
then $\langle E,-\, |\, E,+\rangle=0$ and
$$
\langle E,+\, |\, E',+\rangle=\delta(E-E')=
\langle E,-\, |\, E',-\rangle\, .
$$
In addition to particle scattering states, it is interesting to look for the $\mathfrak{osp}(1|2)$ analog of the $\mathfrak{sp}(2)$ spherical vector. Indeed recall that the spherical vector for the metaplectic representation of $Sp(2,{\mathbb R})$ is the state 
with minimal eigenvalue of 
 the generator $H+K$ of the maximal compact subgroup $SO(2)$. Indeed, this is none other than the harmonic oscillator ground state $\psi_0=\exp(-\frac 12 x^2)$. When $E_0=\frac 12 $, this state is annihilated by $S+iQ$.
For the $\mathfrak{osp}(1|2)$ algebra, we therefore search for states in the kernel of $S+iQ$. For bosonic (even) states, the only solution is again $$\psi_0^B=\exp(-\frac 12 x^2)\, ,$$
which is in the Hilbert space $\mathcal H$
so long as $E_0>0$.
For fermionic (odd) states, we must solve
$$ 
\psi'+\frac{2E_0-1}x \psi + x \psi=0
$$
and find
$$
\psi_0^F=\left\{
\begin{array}{cc}\displaystyle
\frac{e^{-\frac12 x^2}}{x^{2E_0-1}}\, , & x>0\, ,\\[3mm]
\displaystyle
-\frac{e^{-\frac12 x^2}}{|x|^{2E_0-1}}\, , & x<0\, .
\end{array}
\right.
$$
The above state is in ${\mathcal H}$ whenever $E_0<1$. Note that strictly speaking, 
for values of the parameter $E_0$ with $0<E_0<1$ a detailed analysis of the self-adjoint extensions of $Q$ is required to decide which combination(s) of the above two states is actually in the kernel of $S+iQ$. 
The above states will play the {\it r\^ole} of highest weights in the next section.

\section{Particle orthosymplectic representation}
\label{sec:XI}

\noindent
It remains to identify the orthosymplectic representations realized by the particle solutions to the deformation equations.

First we compute  the Casimir operator for  the $\mathfrak{sp}(2)$ subalgebra $(H,K,D)$, which reads
\begin{align}
c_{\mathfrak{sp}(2)} = \frac14 D^2 +\frac12 \big\{ H,K\big\} =
\frac1{16}\hh  [iQ,S]\big([iQ,S]-4\big)\, .
\label{eq:casimir}
\end{align}
Using $[F,x]=x(1-2F)$, we here have \begin{align}\label{algae}
\big[iQ,S\big] = \Big[\frac{\partial}{\partial x}\! +\! \frac{2E_0-1}{x}\, F\hh,\hh x\Big] = 1\!-\!(2E_0-1)(2F-1) ,
\end{align}
so once again find
$$
c_{\mathfrak{sp}(2)}({\mathcal B})=\tfrac{E_0(E_0-2)}4\, \mbox{ and }\,
c_{\mathfrak{sp}(2)}({\mathcal F})=\tfrac{(E_0-1)(E_0+1)}4\, ,
$$
and in turn ${\mathcal H}={\mathcal B}\oplus {\mathcal F}$ obeys
\begin{equation*}
\hh c_{\mathfrak{osp}(1|2)}({\mathcal H})=\frac{E_0(E_0-1)}4=\hh c_{\mathfrak{osp}(1|2)}\big({\mathcal S}(E_0)\big)\, .
\end{equation*}

Unitarity requires that the generators
 $\{iQ, S, H, iD, K\}$ are self-adjoint.
 Our deficiency index analysis shows that this holds
 for all $E_0$, modulo the choice of self-adjoint extension for $0<E_0<1$.

To analyze  the $\mathfrak{osp}(1|2)$ content of the model, we can consider an oscillator-like basis for the generators with the reality condition~\eqref{realosp} by employing the map~\eqref{reallinear}. Indeed, calling
$$
A:=\frac{S+iQ}{\sqrt{2}}\, ,\qquad A^\dagger = \frac{S-iQ}{\sqrt 2}\, ,
$$ 
we have (using~\eqref{algae})
$$
[A,A^\dagger]=
1-(2E_0-1)(2F-1)\, ,
 $$
and $S^+=A^\dagger$, $S^-=A$ obey the 
$\mathfrak{osp}(1|2)$ algebra~\eqref{osp}. 
(Note that this is a different solution to that given in Eq.~\eqref{solution}.)
At this point the operators $A$ and $A^\dagger$ obey the same algebra as analyzed for the oscillator models in Section~\ref{sec:VII}, so we can inherit that analysis, however some care is required when $0<E_0<1$.

Firstly when $|E_0-\frac 12|\geq 1/2$ the self-adjoint extension problem gives a unique answer and indeed there is a unique highest weight state
$$
|E_0,0\rangle=
\left\{
\begin{array}{cc}
\psi_0^B\, ,& E_0\geq1\, ,\\[2mm]
\psi_0^F\, ,& E_0\leq0\, .
\end{array}
\right.
$$  
The descendants of $|E_0,0\rangle$ (generated by acting with $A^\dagger$)
then span the irreducible representation ${\mathcal S}(E_0)$.

When $0<E_0<1$ there are potentially two highest weight states $\psi_0^B$ and $\psi_0^F$, however, we {\it conjecture} that only one combination of these is a zero mode of $A$ for a given choice of self-adjoint extension of $Q$.

As an example consider the undeformed models with $E_0=\frac12$ and $Q=\frac{d}{dx}$. Here, the Hilbert space is ${\mathcal H}=L^2({\mathbb R}^+)\oplus L^2({\mathbb R}^+)$. There are of course no self-adjoint extensions 
of $\frac{d}{dx}$ on the half-line, but it is easy to find one for~$\frac{d}{dx}$ defined on two copies of ${\mathbb R}^+$, namely by viewing pairs of wavefunctions there as the even and odd parts of wavefunctions in $L^2({\mathbb R})$, on which $\frac{d}{dx}$ is essentially self-adjoint. In that case $A\psi^B_0=0$ because~$\psi^B_0$ is the usual harmonic oscillator ground state, while $A \psi^F_0(x)=2\delta(x)\neq0$. The descendants of $\psi_0^B$ then give the unitary irreducible orthosymplectic representation ${\mathcal S}(\frac12)$.

 We have summarized 
 the orthosymplectic representations realized by deformations of superconformal quantum mechanics  in the diagram below:

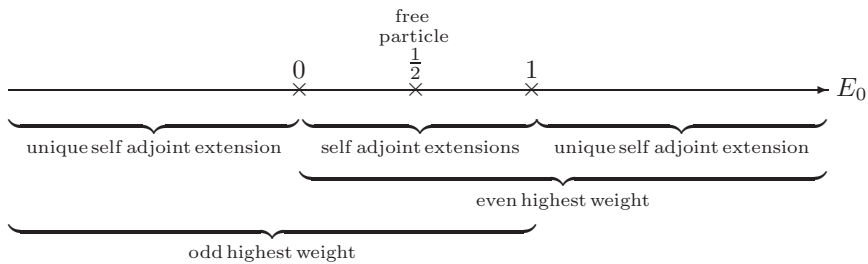
\begin{widetext}

\vspace{1cm}

\begin{center}
\begin{picture}(100,10)(-60,0)

\put(-160,5){\vector(1,0){310}}
\put(-54,2.62){$\times$}

\put(-10,2.62){$\times$}

\put(34,2.62){$\times$}

\put(153,2.62){$E_0$}

\put(-52.8,9.62){$0$}

\put(-9.4,12.62){$\frac12$}

\put(-23,26.82){$\begin{array}{c}\rm \scriptstyle free\, \\[-1.3mm] \rm
\scriptstyle particle\end{array}$}

\put(35.3,9.62){$1$}

\put(-160,-5){$
\underbrace{\hspace{3.85cm}}_{\rm unique\,  self\, adjoint\, extension}
\underbrace{\hspace{3.12cm}}_{\rm self\, adjoint\, extensions}\hspace{-.73mm}\underbrace{\hspace{3.85cm}}_{\rm unique\,  self\, adjoint\, extension}
$}

\put(-50,-25){$\underbrace{\hspace{7cm}}_{\rm even\,  highest\,  weight}$}

\put(-160,-45){$\underbrace{\hspace{7cm}}_{\rm odd\,  highest\,  weight}$}


\end{picture}
\end{center}

\vspace{2.5cm}

\end{widetext}

\section{Summary and Conclusions}

\noindent
Although supersymmetric quantum mechanics has a long history~\footnote{See, for example~\cite{Witten,vanH,Rit}.}, its presence in even the simplest of quantum mechanical models is often underappreciated---both the free particle and harmonic oscillator enjoy a hidden $\mathfrak{osp}(1|2)$
superconformal symmetry realized by employing wave-function parity for  the  Bose--Fermi ${\mathbb Z}_2$-grading.
Given a particle/oscillator Hilbert space, we studied the natural question whether other 
sets of operators realize this algebra.
In higher dimensions the moduli space of such  operators has a particularly interesting geometric structure: For example, on {\it any} (pseudo)-Riemannian manifold whose metric $g_{\mu\nu}$ is the gradient of a covector $g_{\mu\nu}=\nabla_\mu \xi_\nu$, the triplet of operators $\{\xi_\mu\xi^\mu,\, \xi^\mu\nabla_\mu, \, \nabla^\mu \nabla_\mu\}$ generate the algebra $\mathfrak{sp}(2)$.
Including spinors and the Dirac operator, this algebra can be extended to the $\mathfrak{osp}(1|2)$ superalgebra studied here and indeed our study is the special case when the underlying manifold is one-dimensional.
The fact that we were able to give a detailed classification of this space of operators in a  one-dimensional setting suggests that similar general results ought be obtainable in higher dimensions. This is exciting because of its relevance to interacting higher spin and quantum gravity models~\cite{BarsRey, BonezziO,Bonezzi,BonezziDirac,BarsReview}.


The one dimensional solutions to the $\mathfrak{osp}(1|2)$ operator question are parameterized by a one (complex) parameter moduli space.
It would be interesting to try and mimic these results for higher hidden quantum mechanical SUSY algebras, the results of~\cite{P4} indicates that
 this ought be possible~\footnote{Another avenue for generating such model is to study the Dirac equation in more general backgrounds, such as dyonic ones~\cite{dH3}.}. Here, once one studies Hilbert spaces for mechanics in higher dimensions, one expects a moduli space of solutions with more constraining geometric structures than  conformal geometries.
 
 One might wonder whether our results contravene the Stone--Von Neumann theorem on unitary equivalence of Heisenberg representations. This is not the case because  the Plyushchay-type models generate $\mathfrak{osp}(1|2)$ representations with differing values of $E_0$ and inner product by modifying the commutation relation $[a,a^\dagger]=1$  to $[S,S^\dagger]=1-(2E_0-1)(2F-1)$, where $E_0=1/2$ gives the standard harmonic oscillator model. It interesting to note that that this deformation is important for deformations higher spin algebras leading to interactions~\cite{PV,Top2}.

The $E_0=1/2$ orthosymplectic representation is a sum of two discrete series $\mathfrak{sl}(2,\mathbb R)$ representations analogous to the double cover  half integer spin representations in the theory of angular momentum. Indeed, this is the  so-called metaplectic representation of $Sl(2,\mathbb R)$. 
It would be interesting to exponentiate these realizations of $\mathfrak{osp}(1|2)$ representations to give  analogs of the metaplectic representation.


\begin{acknowledgments}
\noindent
We thank Itzhak Bars, Yoon Seok Chae, Jan Derezi\'nski, 
Eric D'Hoker, Rita Fioresi, Rod Gover, Maxim Grigoriev, Bruno Nachtergaele,
Mikhail Plyushchay and Andrea Sacchetti
for discussions and suggestions.
 A.W. thanks
 the Department of Mathematics of the University of Bologna for hospitality.
The work of R.B. was supported by a PDR ``Gravity and extensions''  F.R.S.-FNRS (Belgium) grant.
A.W. was supported in part by a 
Simons Foundation Collaboration Grant for Mathematicians ID 317562.
\end{acknowledgments}

\appendix

\section{General Parity Solutions}
\label{E's favorite appendix}

\noindent
To show that the solution~\eqref{solution} is general, we must
relax the requirement that the ${\mathbb Z}_2$ gradings of the  $\mathfrak{osp}(1|2)$ Lie superalgebra and the Hilbert space ${\mathcal H}$ 
are coincident. Thus we study a general  version of the ansatz Eq.~\eqref{ansatz1}, namely 
$$
S^-=\frac{\partial}{\partial z} 
+ \alpha_+(z) + \alpha_-(z) +(-1)^F\big[\beta_+(z)+\beta_-(z)\big]\, .
$$
Here and in what follows, we denote even/odd functions of $z$ by a subscript~$\pm$.
The second, nonlinear equation in~\eqref{eom} now yields a Dirac-like equation
$$
\big(z\frac{\partial}{\partial z} + 1\big)
\begin{pmatrix}
\beta_+(z)\\ \beta_-(z)
\end{pmatrix}
-2z\alpha_+(z)
\begin{pmatrix}
\beta_-(z)\\
\beta_+(z)
\end{pmatrix}=0\, .
$$
Notice that $\alpha_-(z)$ is completely free while we can solve for $\beta(z)=\beta_+(z)+\beta_-(z)$ in terms of $\alpha_+(z)$ as
$$
\beta(z) = \frac{E_0-\frac12}{z}\hh \exp\Big(2\int^z \alpha_+\Big)\, . 
$$
Hence we find
$$
S^-=\frac\partial{\partial z} + \alpha(z) + (-)^F \hh \, \frac{E_0-\frac12}{z} \exp\Big(2\int^z \alpha_+\Big)\, .
$$
Here $\alpha(z) = \alpha_+(z)+\alpha_- (z)$
and we must set $\alpha_-(z)
 =\tfrac{2E_0-1}{2z}  + a_-(z)$ (with $a_-(z)$ odd and analytic) to cancel the~$1/z$ pole in $S^-$.
 Again, evenness of  $U(z)=\exp\big(-\int^z a_-(z)\big)$
 allows use to gauge away $a_-(z)$.
 This yields
 \begin{multline}
S^-=\frac\partial{\partial z} +
\frac{2E_0-1}z\, F\\ + \alpha_+(z)
+ (-1)^F \, \frac{E_0-\frac12}{z} \Big[\exp\Big(2\int^z \alpha_+\Big)-1\Big]\, ,\nonumber
\end{multline}
which is the sum of our previous $\mathfrak{osp}(1|2)$ odd solution and a mixed $\mathfrak{osp}(1|2)$ parity solution parameterized by the even, analytic function $\alpha_+(z)$. 

The Hamiltonian for this class of models is given by
$$
H=N+E_0+z\alpha_+(z)\, .
$$
The Casimir 
is
again $c_{{\mathfrak{osp}}(1|2)}=E_0(E_0-1)/4$
which suggests that this solution is gauge equivalent to our previous one.
Indeed the additional gauge transformation $U(z)=\exp\big(-\int^z a_+(z)\big)$
can be used to remove the $a_+(z)$ dependence of the Hamiltonian and the
ladder operator $S^-$, whence
$
H=N+E_0$ and
$S^-=\frac\partial{\partial z} +
\frac{2E_0-1}{z} F$.
Remembering that $S^+=z$, we recognize our previous solution in Eq.~\eqref{solution}.


\section{Orthosymplectic representation theory}
\label{repth}
The following material reviews basic results from the representation theory of $\mathfrak{sl}(2,\mathbb R)$ and $\mathfrak{osp}(1|2)$.
We also provide a translation between  common notations found in the literature and those used here.

The Lie algebra $\mathfrak{sl}(2,\mathbb C)=\{e,h,f\}$ where~\footnote{The conventions $X_+=e$, $H=h/2$ and $X_-=-f$ for which
$
[H,X_+]=X_+$,
$[X_+,X_-]=-2H$,
$[X_-,H]=X_-
$, 
and $X_0^\dagger=D_0$, $X_\pm^\dagger=X_\mp$ are also common.
}
\begin{equation}\label{sl2}
[h,e]=2\hh e\, ,\quad [e,f]=h\, ,\quad [f,h]=2f\, ,
\end{equation}
has two inequivalent real forms; since we are interested in quantum mechanical models with infinite dimensional Hilbert spaces, our focus is on the non-compact $\mathfrak{sl}(2,\mathbb R)\cong \mathfrak{sp}(2,\mathbb R)$ form~\footnote{The congruence between $\mathfrak{sl}(2,{\mathbb R})$ and the worldline conformal algebra $\mathfrak{so}(2,1)=\{J_0,J_1,J_2\}$ is often useful. This is given by $J_0=h/2=J_0^\dagger$, $J_1=\frac i2 (e+f)=J_1^\dagger$ and $J_2=\frac12 (e-f)=J_2^\dagger$.}
\begin{equation}
\label{real}
e^\dagger = -f\, ,\quad h^\dagger = h\, ,\quad f^\dagger = -e\, .
\end{equation}
For example, the harmonic oscillator obeys the above by setting $h=H=a^\dagger a+\frac 12$, $e=\frac 12(a^\dagger)^2$ and $f=-\frac12\hh  a^2$.
The real linear map
\begin{equation}
\label{realtate}
e\mapsto \tfrac12(h+e-f)\, \quad
h\mapsto -e-f\, ,\quad f\mapsto \tfrac12(h-e+f)\, ,
\end{equation}
preserves the $\mathfrak{sl}(2)$ Lie algebra but gives reality conditions
\begin{equation}\label{fp}
e^\dagger = e\, ,\quad h^\dagger=-h\, , \quad f^\dagger=f\, .
\end{equation}
This choice of $\mathfrak{sl}(2,\mathbb R)$
generators corresponds to the free particle on a line with $e=\frac12 x^2$\, , $h=x\frac{\partial}{\partial x}+\frac12$ and $f=H=-\frac 12 \frac{\partial^2}{\partial x^2}$.
 
 \medskip
 
The Lie algebra~\eqref{sl2} is extended to the ${\mathbb Z}_2$ graded algebra $\mathfrak{osp}(1|2)\cong \mathfrak{sp}(2)\lpl {\mathbb R}^2$ by adding odd generators $s$ and $q$ that obey
\begin{equation}
\label{relations}
\{s,s\}=  e\, ,\quad \{s,q\}=\frac 12\hh  h\, ,\quad
\{q,q\}=- f\, .
\end{equation}
In the notation of the introduction, $s=\frac 12S^+$,  $q=\frac12 S^-$ so the remaining commutation relations may be read off the second line of~\eqref{osp} and read
$$
[s,f]=q\, ,\quad
[h,s]=s\, ,\quad
[q,h]=q\, ,\quad
[q,e]=s\, .
$$

 Given the reality conditions~\eqref{real}, there are two inequivalent  reality conditions for the odd generators~\cite{Rittenberg}
\begin{equation}\label{realosp}
s^\dagger=\pm q\, ,\quad q^\dagger=\pm s\, .
\end{equation}
The first choice above  is realized by the harmonic oscillator with $s=\frac12 a^\dagger$ and $q=\frac 12a$.
The real linear map
\begin{equation}\label{reallinear}
s\mapsto \tfrac{1}{\sqrt{2}} (s+q)\, ,\quad
q \mapsto \tfrac{1}{\sqrt{2}} (-s+q)\, , \end{equation}
induces the map~\eqref{realtate}
through the relations~\eqref{relations} and preserves the $\mathfrak{osp}(1|2)$ algebra.
It gives again the free particle-type reality conditions~\eqref{fp} and reality conditions
$$
s^\dagger = \pm s\, , \quad q^\dagger =\mp q\, .
$$
The first case corresponds to a free particle on the line with $s=\frac 12 x$ and $q=
\frac i2 Q=\frac 12\frac{ \partial}{\partial x}$.

Unitary irreducible representations of $\mathfrak{sl}(2,\mathbb R)$ are infinite dimensional~\cite{LANG} and fall into three series: principal, supplementary and discrete.
Unitary irreducible representations of $\mathfrak{osp}(1|2)$ are also infinite dimensional and are built from a direct sum of discrete series representations~\cite{Furutsu}: Call
 \begin{widetext}
$$
{\mathcal D}(E_0):={\rm span}\big\{|E_0,2k\rangle =e^k|E_0,0\rangle\hh \big |\hh  k\in {\mathbb Z}_{\geq 0}\, ,\, h|E_0,0\rangle=E_0|E_0,0 \rangle\, , f|E_0,0\rangle=0
\big\}\, .
$$
\end{widetext}
The reality conditions~\eqref{realosp} imply that
$$
\big|\big||E_0,2k\rangle\big|\big|^2= k! E_0(E_0+1)\cdots (E_0+k-1) \big|\big||E_0,0\rangle\big|\big|^2\, .
$$
The right hand side above is certainly positive 
whenever the ``ground state energy'' $E_0\in {\mathbb R}_{>0}$. Indeed the Hilbert space ${\mathcal D}(E_0)$ for real positive $E_0$ is the unitary irreducible (positive) discrete series representation of $\mathfrak{sl}(2,\mathbb R)$. It has quadratic Casimir 
$$
c_{\mathfrak{sp}(2)}=\frac14\hh  h^2 + \frac12 \hh (ef+fe)
$$
given by 
$$
c_{\mathfrak{sp}(2)}\big({\mathcal D}(E_0)\big)
= \frac{E_0(E_0-2)}4=\frac14 \big[(E_0-1)^2-1\big]\, .
$$
Hence the representations ${\mathcal D}(E_0)$ and ${\mathcal D}(2-E_0)$ have the same Casimir. In particular, the harmonic oscillator Hilbert space is
$$
{\mathcal D}(1/2)\oplus
{\mathcal D}(3/2) \, ,
$$
where both discrete series representations have $c_{\mathfrak{sp}(2)}=-\frac 3{16}$. Indeed ${\mathcal D}(1/2)$ is spanned by even number operator eigenstates $\{|0\rangle,|2\rangle,\ldots\}$ with $|1/2,0\rangle=|0\rangle$ while 
${\mathcal D}(3/2)$ is spanned by odd  eigenstates $\{|1\rangle,|3\rangle,\ldots\}$ with $|3/2,0\rangle=|1\rangle$. 
The above Hilbert space also forms the metaplectic representation of the group $Sl(2,{\mathbb R})$; this can be viewed as the non-compact analog of the double cover spin representations of $SU(2)$.

The unitary irreducible representations of $\mathfrak{osp}(1|2)$ generalize the harmonic oscillator example and are given by the ${\mathbb Z}_2$-graded vector space~\cite{Furutsu} 
\begin{widetext}
\begin{equation*}
\begin{split}
{\mathcal S}(E_0)&={\rm span}\Big\{|E_0,2k\rangle =e^k|E_0,0\rangle,
|E_0+1,2k\rangle =e^ks|E_0,0\rangle
\hh \big |\hh  k\in {\mathbb Z}_{\geq 0}\, ,\, h|E_0,0\rangle=E_0|E_0,0 \rangle\, , f|E_0,0\rangle=0=
q|E_0,0\rangle
\Big\}\\[1mm]
&=
{\mathcal D}(E_0)\oplus 
{\mathcal D}(E_0+1)=
{\mathcal D}(E_0)\oplus 
s{\mathcal D}(E_0)\, .
\end{split}
\end{equation*}
\end{widetext}
where $E_0>0$. The respective $\mathfrak{sp}(2)$ Casimirs differ by~$\frac12\big(E_0-\frac12\big)$. The $\mathfrak{osp}(1|2)$ Casimir is
\begin{equation}\label{ospCas}
c_{\mathfrak{osp}(1|2)}=c_{\mathfrak{sp}(2)}+
\frac12\hh\big(qs-sq)\, .
\end{equation}
This can be reexpressed in the enveloping algebra as
$$
c_{\mathfrak{osp}(1|2)}=[q,s]\big([q,s]-\tfrac12\big)\, .
$$
On the orthosymplectic discrete series it
takes the value
$$
c_{\mathfrak{osp}(1|2)}\big({\mathcal S}(E_0)\big
)
=\frac14\hh E_0(E_0-1)=\frac14 \hh \big[(E_0-\tfrac12)^2 - \tfrac14\big]\, .
$$
Observe that this is minimized by $E_0=\frac 12$ which corresponds to the harmonic oscillator.



\begin{thebibliography}{99}

\bibitem{P1}
M. S. Plyushchay,
``Supersymmetry without fermions'',
[hep-th/9404081].


  
 \bibitem{BarsRey} 
  I.~Bars,
  Phys.\ Rev.\ D {\bf 64}, 126001 (2001)
  [hep-th/0106013]; I. Bars and S.-J.~Rey, Phys.Rev. {\bf D64} (2001) 046005 [hep-th/0104135]; 

 \bibitem{BonezziO} R. Bonezzi, E. Latini, A. Waldron,
Phys. Rev. {\bf D82} (2010) 064037,
[arXiv:1007.1724 [hep-th]]. 


\bibitem{Bonezzi} 
  R.~Bonezzi, O.~Corradini and A.~Waldron,
  Phys.\ Rev.\ D {\bf 90}, no. 8, 084018 (2014)
  [arXiv:1407.5977 [hep-th]].

\bibitem{FG} C. Fefferman and C.R. Graham,
\'Elie Cartan et
les Mathematiques d'Aujourd'hui (Ast\'erisque, 1985) 95.

\bibitem{BonezziDirac} 
  R.~Bonezzi, O.~Corradini, E.~Latini and A.~Waldron,
  Phys.\ Rev.\ D {\bf 91}, no. 12, 121501 (2015)
  [arXiv:1505.01013 [hep-th]].

\bibitem{BarsReview}
I. Bars,
 Class. Quant. Grav. {\bf 18} (2001) 3113,
(hep-th/0008164).

\bibitem{dAFF} 
  V.~de Alfaro, S.~Fubini and G.~Furlan,
  Nuovo Cim.\ A {\bf 34} (1976), 569.

\bibitem{dH1}
E.~D'Hoker and L.~Vinet,
 Phys.\ Lett.\  {\bf 137B}, (1984) 72 (1984).

\bibitem{Wigner}
E. P. Wigner,
Phys. Rev. {\bf 77} (1950) 711.

\bibitem{Yang-Dunkl}
L. M. Yang, Phys. Rev. {\bf 84} (1951) 788; C. F. Dunkl, Trans. Amer. Math. Soc. {\bf 311} (1989) 167.


\bibitem{P3}
M. S. Plyushchay,
Mod. Phys. Lett. A{\bf 11} (1996) 397,
(hep-th/9601141).


\bibitem{TheTop}
I.E. Cunha, N.L. Holanda, F. Toppan,
Phys.Rev. D{\bf 96} (2017) 065014 
(arXiv:1610.07205 [hep-th]).


\bibitem{Rittenberg}
M. Scheunert,  W. Nahm, and  V. Rittenberg, 
J. Math. Phys.
{\bf 18} (1977)  146;
J. Math. Phys.  {\bf 18} (1977), 155.


\bibitem{LANG} See, for example, S. Lang, ``$Sl_2(\mathbb R)$'', Graduate Texts in Mathematics {\bf 105}, Springer-Verlag New York, 2nd Ed. 1988, for a complete account of the representation theory of $Sl_2(\mathbb R)$.

\bibitem{Furutsu}
H. Furutsu and T. Hirai, 
J. Math. Kyoto Univ. {\bf 28} (1988), 695.
See also A. El Gradechi, ``
Supercoherent states and geometric quantization and of a super K\"ahler supermanifold'', in Quantization and Infinite Dimensional Systems, 
 I.P. Antoine et al, Plenum Press 1994; 
 I. Bars and M. G\"unaydin, J. Math. Phys. {\bf 20} (1979) 1977.



\bibitem{P2}
 M.S. Plyushchay,
Ann. Phys. {\bf 245} (1996) 339,
(hep-th/9601116).

\bibitem{dH2}
E.~D'Hoker and L.~Vinet,
 Commun.\ Math.\ Phys.\  {\bf 97} (1985) 391.


\bibitem{Reed} There is a large literature on the theory of self-adjoint extensions, a useful 
starting point is: M. Reed and B. Simon, ``Methods of Modern Mathematical Physics'', vol. I and II, Academic Press, 1975.

\bibitem{Derezinski}
J.  Derezi\'nski and S. Richard,
Ann. Henri Poincar\'e  {\bf 18} (2017), 
869.

\bibitem{Witten}
E. Witten, 
Nucl. Phys. B{\bf 188} (1981) 513.

\bibitem{Rit}
M. de Crombrugghe and V. Rittenberg, 
Ann. of Phys.
{\bf 151} (1983)
99. 

\bibitem{vanH}
J.W. Van Holten and 
P. Salomonson
Nucl. Phys. B{\bf 196} (1982) 509.


\bibitem{P4}
J. Gamboa, M.S. Plyushchay and J. Zanelli,
Nucl.Phys. B{\bf 543} (1999) 447,
(hep-th/9808062);
M.S. Plyushchay,
Int.J.Mod.Phys. A{\bf 15} (2000) 3679, (hep-th/9903130);
 F. Correa, M. A. del Olmo, M. S. Plyushchay, 
 Phys. Lett. B{\bf 628} (2005) 157
(hep-th/0508223)
10.1016/j.physletb.2005.09.046.
Phys.Lett. B628 (2005) 157-164.
F. Correa and  M. S. Plyushchay,
Ann. Phys. {\bf 322} (2007) 2493,
(hep-th/0605104);
10.1016/j.aop.2006.12.002.
Annals Phys. 322 (2007) 2493-2500;
P. A Horvathy, M. S. Plyushchay, and M. Valenzuela, Phys.Rev. D{\bf 77} (2008) 025017,
(arXiv:0710.1394 [hep-th]).


\bibitem{dH3}
E.~D'Hoker and L.~Vinet,
 Lett.\ Math.\ Phys.\  {\bf 12} (1986) 71;
 Phys.\ Rev.\ Lett.\  {\bf 55} (1985) 1043.

\bibitem{PV}
S.~F.~Prokushkin and M.~A.~Vasiliev,
  Nucl.\ Phys.\ B {\bf 545} (1999) 385,
  (hep-th/9806236).
  
\bibitem{Top2}  
F. Toppan and M. Valenzuela,
``Higher Spin Symmetries and Deformed Schr\"odinger Algebra in Conformal Mechanics'',
arXiv:1705.04004 [hep-th].


%
%
%
%
%
%
%
%
%
%
%
%
%
%
%
%
%
%
%
%
%
%
%
%
%
%
%
%
%
%
%
%
%
%
%
%
%
%
%
%
%
%
%
%
%
%
%
%
%
%
%
%
%
%
%
%
%
%
%
%
%
%
%
%
%
%
%
%
%
%
%
%
%
%
%
%
%
%
%
%
%





 \end{thebibliography}
\end{document}